\documentclass[runningheads]{llncs}
\usepackage[T1]{fontenc}
\usepackage{url}
\usepackage{booktabs}

\usepackage{graphicx}
\usepackage{cite}
\usepackage{amsmath,amssymb,amsfonts}
\usepackage{graphicx}
\usepackage{textcomp}
\usepackage{xcolor}
\usepackage{hyperref}
\usepackage{algorithm}
\usepackage{algpseudocode}
\usepackage{subcaption}

\makeatletter
\newcommand\footnoteref[1]
{\protected@xdef\@thefnmark{\ref{#1}}\@footnotemark}
\makeatother
\begin{document}
%
\title{From Online Behaviours to Images:\\A Novel Approach to Social Bot Detection\thanks{This work was partially supported by project SERICS (PE00000014) under the NRRP MUR program funded by the EU - NGEU; by the Integrated Activity Project TOFFEe (TOols for Fighting FakEs) \url{https://toffee.imtlucca.it/}; by the IIT-CNR funded Project re-DESIRE (DissEmination of ScIentific REsults 2.0); by `Prebunking: predicting and mitigating coordinated inauthentic behaviors in social media' project, funded by Sapienza University of Rome.}}
\titlerunning{A Novel Approach to Social Bot Detection}
%
\author{Edoardo Di Paolo\inst{1}\orcidID{0000-0001-9216-8430} \and
Marinella Petrocchi\inst{2}\orcidID{0000-0003-0591-877X} \and
Angelo Spognardi\inst{1}\orcidID{0000-0001-6935-0701}}

%
\institute{
Computer Science dept., Sapienza University of Rome, Italy
\email{\{dipaolo,spognardi\}@di.uniroma1.it}\\
 \and
IIT-CNR, Pisa, Italy\\
\email{marinella.petrocchi@iit.cnr.it}}
\maketitle              
%


\begin{abstract}
    Online Social Networks have revolutionized how we consume and share information, but they have also led to a proliferation of content not always reliable and accurate.
    One particular type of social accounts is known to promote unreputable content, hyperpartisan, and propagandistic information. 
    They are automated accounts, commonly called bots. Focusing on Twitter accounts, 
    we propose a novel approach to bot detection: we first propose a new algorithm that transforms the sequence of actions that an account performs into an image; then, we leverage the strength of Convolutional Neural Networks to proceed with image classification. 
    We compare our performances with state-of-the-art results for bot detection on genuine accounts / bot accounts datasets well known in the literature. The results confirm the effectiveness of the proposal, because the detection capability is on par with the state of the art, if not better in some cases. 
\end{abstract}

\section{Introduction}


With the advent of the internet and Online Social Networks (OSNs), production and fruition of information feature less mediated procedures, where content and quality do not always go through a rigorous editorial process~\cite{gangware2019weapons,ceron2015internet,valkenburg2013comm}. Thus, although OSNs make our lives easier by giving us immediate access to information and allowing us to exchange opinions about anything, the danger of being exposed to false or misleading news is high~\cite{fakenewsgeneral, misinformationtrends, healthmisinformation}.
The promotion of disinformation on OSNs has often been juxtaposed with the existence of automated accounts known as bots. As an example, Shao et al., in~\cite{shao2018anatomy}, have highlighted the role of Twitter bots, showing how bots were primarily responsible for the early spread of disinformation, interacting with influential accounts through mentions and replies.



The struggle between bots hunters and bots creators has been going on for many years now~\cite{cresci2020decade}, and the actions of these automated accounts with malicious intent have influenced even the purchase of Twitter itself -just remember the \$44 billion deal that went up in smoke precisely because of concerns about the unquantified presence of bots on the platform\footnote{Elon Musk terminates \$44B Twitter deal. Online: \url{https://nypost.com/2022/07/08/elon-musk-terminates-44-billion-twitter-deal/} August 8, 2022.}.



In this study, we provide a novel approach to bot detection, by leveraging the remarkable advancements in the field of image recognition~\cite{Krizhevsky17ImageNet,Liu22ConnNet}. To the best of the authors' knowledge, no methodology or tool so far defined for bot detection leverage image recognition.  
In particular, we will  exploit the potential of Convolutional Neural Networks (CNNs) to classify Twitter accounts as bots or not. 

Based on the premise that automated accounts are often programmed to carry out spam and/or disinformation campaigns, numerous works in the literature have proposed detection approaches that leverage coordination and synchronism features of accounts, see, e.g.~\cite{yu2015,Cao:2014,Cresci2018fingerprinting}. The intuition is that the online activities of a group of automated accounts -all devoted carrying out a certain strategy- are more similar to each other than those of genuine accounts. 
This is the leit motif from which the concept of {\it Digital DNA}, originally introduced by Cresci et al. in~\cite{DBLP:journals/expert/CresciPPST16}, and the detection technique known as {\it Social Fingerprinting}~\cite{cresci2017} came to life. 
The digital DNA of an online account represents the sequence of actions of that account and each action is associated to a symbol from a pre-defined alphabet. By associating each symbol with a color, it is possible to transform the sequence into an image, where each pixel represents the color of the corresponding symbol. The assumption that leads us to exploit image classification to perform bot detection is that images of bot accounts are similar to each other, and different from those of genuine accounts, given the different behavior of the two categories of accounts. 

We thus propose an algorithm to transform sequences of digital DNA into images and we run pre-trained CNNs, such as VGG16, ResNet50, and WideResNet50~\cite{ResNet,simonyan2014very}  over the generated images. DNA sequences are extracted from Twitter accounts, both genuine and bot, of public datasets well-known in the bot detection literature. Where accounts' timelines are too short to produce good quality images, we have enhanced the latter, even turning features of the accounts other than Digital DNA into part of it.

\paragraph{Main Contributions:} The main contributions of this work are as follows:
\begin{itemize}
    \item Definition and implementation of a new approach to bot detection, based on image recognition;
\item Validation of the approach by comparing our performance with state-of-the-art performances on publicly-released datasets: bot detection via image recognition achieves the same perfomances as obtained in the literature, when not better. 
\end{itemize}

We argue that the investigations here presented make the transition from sequence of actions to sequence of pixels for bots detection look promising. Of course, the literature is filled with more than good work in the field. Still, we find the approach itself interesting because it leverages well-established image recognition techniques. Thus, a way forward for further experimentation.

\section{Related Work}
Bot detection is a topic that began to be studied more than 10 years ago, when social networks became increasingly popular, and interests in spamming, spreading propaganda content, and increasing one's notoriety on the platforms grew tremendously. 
Different techniques have followed over the years, from classifying via traditional machine learning exploiting features in the accounts profile -- the very first attempts in this direction are the papers by Mustafaraj and Metaxas~\cite{mustarafi10} and Yardi et al.~\cite{yardi2010detecting}, both dated 2010 --, to using deep learning techniques. 
We therefore feel it is appropriate to list some of the work on the subject, without however intending to propose an exhaustive list. 


\paragraph{Traditional machine learning approaches.}

 Botometer is probably the most well-known tools in the literature for bot unveiling~\cite{FerraraArming2019}; 
it 
is based on a supervised machine learning approach employing Random Forest classifiers. 
The last version, Botometer v4, has been recently shown to perform well for
detecting both single-acting bots and coordinated campaigns~\cite{DBLP:conf/cikm/Sayyadiharikandeh20}. v4 provides a useful lite version, BotometerLite\footnote{\url{https://cnets.indiana.edu/blog/2020/09/01/botometer-v4/}},  which does not interface with Twitter, but simply takes the tweet, retrieves the author, and does the necessary follow-up analysis. This light version only needs the information in the user profile to perform bot detection and hence can also process historical data published by accounts that are no longer active.

Over the years, there has been no limit in engineering accounts' features, to feed them to machine learning classifiers, e.g., the length of usernames, the reposting rate, some temporal patterns and the similarity of message contents based on Levenshtein distance~\cite{efthimion2018supervised}, just to name a few.



\paragraph{Deep learning approaches.}
Hayawi \textit{et al.} in~\cite{deeprobot} propose  DeeProBot, where only some of the user profile features (the username’s length and the number of followers) are exploited in order to classify single accounts with an LSTM (Long Short-Term Memory) network. 
Najari \textit{et al.} in~\cite{ganbot} use a GAN (Generative Adversarial Network) associated with a LSTM network. GANs generate bot samples to obtain more information about their behavior. 
RoSGAS (Reinforced and Self-supervised GNN Architecture Search)~\cite{rosgas} is based on multi-agent deep reinforcement learning. 

In~\cite{nlpbotdetect}, the authors propose a deep neural network based on a LSTM architecture processing the texts of tweets, thus exploiting NLP techniques. Their intuition is that bot accounts produce similar contents; therefore, analyzing texts can help the classification. 
Authors of~\cite{nlpwordembeddings} propose a text-based approach using a bidirectional LSTM.
Work in~\cite{dabot} presents a framework with deep neural networks and active learning to detect bots on Sina Weibo.

All of the cited works have been tested on publicly released bot datasets and achieve very good performances (greater than 0.9), considering standard classification metrics such as accuracy, precision, recall and Area Under the Curve.
\paragraph{Graph-based approaches.}
Detection techniques also take into account  graph neural networks, where the social network is seen as a graph, where users are the nodes and the edge between two nodes represents a relationship between users, such as, e.g.,  a followship or retweet relationship. Thus, features derived from the social graph were considered along with profile and timeline fatures to train new models.  An example is the work by Alhosseini~\textit{et al.}~\cite{detectme} which achieves very high perfomances, still on publicy released datasets~\cite{graphDataset}. 

\paragraph{Behavioral analysis.}

Approximately from 2014, a number of research teams, independently,
proposed new approaches
for detecting coordinated behavior
of automated malicious accounts, see, e.g.,~\cite{IntSys2015, yu2015}. That line of research does not consider individual account properties, but rather properties in common with a group of accounts,  like  detection of loosely synchronized actions~\cite{Cao:2014}.
It is precisely in the context of the analysis of synchronism and coordination of the account behaviours that the idea of associating symbols with account actions arose, so that the timeline is represented as a string, so called Digital DNA~\cite{DBLP:journals/expert/CresciPPST16}. The concept of Digital DNA is fundamental in the present work and will be introduced in the next section. Recently, Digital DNA has been re-analysed by Gilmary et al. in~\cite{GilmaryDAN2022}, where they measure the entropy of the DNA sequences. Even in this case, the detection perfomances result in very high values.



This brief roundup of work might lead one to think that bot detection is a solved task. Unfortunately, bots evolve faster than detection methods~\cite{ferrara2016rise,cresci2017}, the latter are not suitable for detecting all kinds of bots~\cite{mazza2022investigating}, and existing datasets for doing training are inherently built according to peculiar accounts characteritics~\cite{olteanu2019social,tan2023botpercent}. 

We therefore conclude this section by pointing out how, perhaps, the task can never be solved in its entirety~\cite{rauchfleisch2020false,cresci2020decade}, and that, since we still have room for investigation, relying on image detection and state-of-the-art tools in this regard seems to us to be a good track to take.

\section{Useful Notions}
\subsection{Digital DNA}\label{sec:digitalDNA}
The biological DNA contains the genetic information of a living being and is represented by a sequence which uses four characters representing the four nucleotide bases: A (\textit{adenine}), C (\textit{cytosine}), G (\textit{guanine}) and T (\textit{thymine}). Digital DNA is the counterpart of biological DNA and it encodes the behaviour of an online account. In particular, it is a sequence consisting of $L$ characters from a predefined alphabet $\mathbb{B}$: 
\begin{equation}
\mathbb{B} = \{\ \sigma_1, \sigma_2, \sigma_3, ..., \sigma_N \}
\label{eq:digitalDNA}
\end{equation}

In \autoref{eq:digitalDNA} each $\sigma$ is a symbol of the alphabet and a digital DNA sequence will be defined as follow:
\begin{equation}
s = (\sigma_1, \sigma_2, ..., \sigma_n), \; \sigma_i \in \mathbb{B} \; \forall \; i = 1, ..., n.
\end{equation}
Each symbol in the sequence denotes a type of action. In the case of Twitter, a basic alphabet is formed by the 3 actions representing the types of tweets:
\begin{equation}
\mathbb{B} = \left\{\begin{array}{c}
        A = \text{tweet}, \\
        C = \text{reply}, \\
        T = \text{retweet}
        \end{array}\right\}
        \label{eq:alphabetExample}
\end{equation}
According to the type of tweets, it is thus possible to encode the account timeline, which could be, e.g., the 
following $s = ACCCTAAACCCCCCTT$. 

Strings of digital DNA were compared to each other in~\cite{Cresci2018fingerprinting}: the longer the \textit{longest common substring} of a group of accounts, the more likely that group is made up of accounts programmed to complete a similar task, i.e., the more likely those accounts are automated accounts.

\subsection{Convolutional Neural Networks}\label{sec:cnn}
Given the recent and noteworthy ~\cite{CNNAdvancements, sultana2018advancements,Krizhevsky17ImageNet,Liu22ConnNet} advancements in the field of Convolutional Neural Networks (CNNs), we asked ourselves whether these networks could be used to classify Twitter accounts into bot / human.

CNNs are typically composed of three layers: convolutional layers, pooling layers, and fully connected layers. The convolutional layer is the fundamental component of a CNN and it requires most of the computation. The input is an image and the dimension of the input image changes depending on whether it is grayscale or colored. 
Combined with the convolutional layer, there is the ``filter'' which is a matrix of small size. From the convolution operation, we have in output a ``filtered'' image which is a sequence of dot products between the input pixels and the filter. Afterward, an activation function can be applied to the output. It is also possible to optimize the convolution layer through some hyperparameters, such as the ``stride'' and the ``padding''. The former represents the amount of movement of the filter over the input, the latter is the process of padding the border of the input.
The second type of layer is the ``pooling'' layer. A pooling layer is used to downsample the given input. There are two main types of pooling: max-pooling and average-pooling. 
The third type of layer is the ``Fully-Connected'' (FC) layer, also known as a ``dense'' layer. The neurons in a FC-layer receive input from all the neurons in the previous layer. 



\section{From Digital DNA to Images}
To the best of our knowledge, no approaches in the literature take advantage of image classification 
to classify social bots. 
The aim is transforming each account’s DNA sequence into an image. Given the similarity in the sequences of bots' actions with respect to those of genuine accounts, the intuition is that an image classifier might work well in the bot detection task. 

 The literature offers some DNA-to-image conversion algorithms~\cite{cgr, dnatoimagesref}. We tried experimenting with these conversion algorithms, but we did not get significant results. Thus, we decided to propose an ad-hoc conversion algorithm, which transforms a digital DNA sequence into a bidimensional object.


\autoref{alg:images_algo} shows the pseudocode for passing from Digital DNA to an image.  Since CNNs expect images of the same size, we first consider the string of maximum length and check whether the length is a perfect square. If not, we consider the  perfect square closest to and strictly largest than the maximum length. By doing so, it is possible to transform all the strings to images of equal size\footnote{As an example, strings as long as 10000 characters are represented by images of size $100$x$100$.}. 

After arbitrarily deciding a RGB color to assign to each symbol in the alphabet, the image is colored pixel by pixel based on the coors assigned to the correspondent symbol. The coloring is done as long as the length of the input string is not exceeded; therefore, if the sequence is not the one with the maximum length, this will result in a black part of the image.
All images created are in grayscale; we tried also with colored images,  but there was no significant improvement in the final results. 

\begin{algorithm}[h]
\small
\caption{From Digital DNA to image: Pseudocode}\label{alg:images_algo}
\hspace*{\algorithmicindent} \textbf{Input:} List of DNA sequences \\
\hspace*{\algorithmicindent} \textbf{Output:} DNA images
\begin{algorithmic}[1]

	\State $n \gets \text{Length of the longest DNA sequence}$
	\If {$n \text{ is a perfect square}$}
		\State $L \gets \sqrt{n}$
	\Else
		\State $L \gets \texttt{get\_closest\_square\_number(n)}$
	\EndIf
	\State $P \gets \text{dict with symbols and colors}$
	
	\For{\texttt{each DNA sequence}}
    	\State $I \gets \texttt{create\_image(width=L, height=L)}$
    	
    	\For{\texttt{row in range(L)}}
            \For{\texttt{col in range(L)}}
            	\State $k \gets \texttt{(row * L) + col}$
            	\If {$k < n$}
            		\State $I[row,\;col] \gets P[DNA[k]]$
            	\EndIf
            \EndFor
        \EndFor
    \EndFor
    
\end{algorithmic}
\end{algorithm}

\section{Datasets}\label{sec:datasets}

This section introduces the datasets on which we tested our detection technique. The datasets are all publicly available.

\subsection{Cresci-2017}
Firstly introduced in~\cite{cresci2017}, this dataset consist of bots and genuine accounts. The kind of bots are various, like  bots engaged in online political discussions, bots promoting specific hashtags, and bots advertising Amazon products. In our study, we evaluated $991$ bots and $1,083$ genuine accounts for a total of $2,074$ samples.

The first step of the procedure is the generation of the DNA sequences for each account in the dataset. We rely on the alphabet in Section~\ref{sec:digitalDNA}(\autoref{eq:alphabetExample}), which considers three symbols, associated to three basic activities on the Twitter platform: Tweet, Retweet, Reply.
After the generation of the DNA strings, we apply the algorithm in~\autoref{alg:images_algo} to generate the images.

\begin{figure}[h!]
\centering
\begin{subfigure}{.5\textwidth}
  \centering
  \includegraphics[width=.5\textwidth]{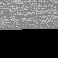}
  \caption{}
  \label{fig:humanc17}
\end{subfigure}%
\begin{subfigure}{.5\textwidth}
  \centering
  \includegraphics[width=.5\textwidth]{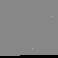}
  \caption{}
  \label{fig:botc17}
\end{subfigure}
\caption{Representation as images of a genuine (left) and bot (right) account belonging to Cresci-2017.}
\label{fig:cresci2017images}
\end{figure}

\autoref{fig:cresci2017images} show two images, representing a genuine and a bot account, \textit{resp.} belonging to the Cresci-2017 dataset. Some noise in \autoref{fig:humanc17} distinguishes this account from that of the bot (\autoref{fig:botc17}). Intuitively, a CNN is able to pick up these differences and, thus, classify the accounts in the correct way. 

\subsection{Cresci-Stock 2018}
First introduced by Cresci et al. in~\cite{Cresci2018FAKEEO}, 
   this dataset 
    consists 
    of both genuine and automated accounts tweeting so-called `cashtags', i.e., specific Twitter hashtags that refer to listed companies. Part of the automated accounts have been found to act in a coordinated fashion,  
in particular by mass retweeting cashtags of low capitalization companies.
In our study, we used $6,842$ bots and $5,882$ genuine users, for a total of $12.724$ labeled samples.

\begin{figure}[t!]
\centering
\begin{subfigure}{.5\textwidth}
  \centering
  \includegraphics[width=.5\textwidth]{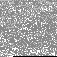}
  \caption{}
  \label{fig:humanc18}
\end{subfigure}%
\begin{subfigure}{.5\textwidth}
  \centering
  \includegraphics[width=.5\textwidth]{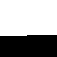}
  \caption{}
  \label{fig:botc18}
\end{subfigure}
\caption{Human and bot accounts on Cresci-Stock-2018 dataset.}
\label{fig:cresci2018images}
\end{figure}

In \autoref{fig:cresci2018images}, it is possible to see the noise which distinguishes a genuine account (\autoref{fig:humanc18}) from a bot account (\autoref{fig:botc18}).

In this case, the image representing the bot is almost completely white, due to the homogeneity of the actions it performes on the social network and due to the choice of the colors used for the different pixels in creating the images. 

\subsection{TwiBot20}
Firstly introduced in~\cite{twibot20}, TwitBot20 is a very large dataset with almost 230k Tiwtter accounts. Of these, the authors provide a total of $11,746$ labeled samples, $6,561$ bots and $5,185$ genuine accounts.  The dataset covers diversified bots and genuine users `to better represent the real-world Twittersphere' and, at the time of the publication of the article presenting it, it represented 
`the largest Twitter bot detection benchmark'.  


In TwitBot20, bot sequences of activities  are very similar to those of genuine accounts. This of course makes the images generated by the sequences similar to each other and there is limited information to highlight specific behavioral patterns.
Furthermore, TwiBot20 features accounts with a maximum of 200 tweets per user; therefore, the images are quite small (15x15). We attempted to enlarge the images, but the results were not good. Details are in~\autoref{sec:exp-res}.

\section{Experiments and Results}\label{sec:exp-res}

For the experiments, we use PyTorch Lightning~\footnote{\url{https://github.com/PyTorchLightning/pytorch-lightning}} to produce readable and reproducible code and it allows to spend less time on engineering the code. We also adopt WanDB~\cite{wandb} to keep track of metrics. Regarding the loss function, we consider the cross entropy (\autoref{eq:cross_entropy_func}).

\begin{equation}
Loss = -(y \log(p) + (1 - y) \log(1-p))
\label{eq:cross_entropy_func}
\end{equation}

The upperbound to the number of epochs is set to 50. However, we use \texttt{EarlyStopping} to monitor the accuracy (or loss) on the validation set: if it the accuracy does to increase ({\it resp.}, the loss does not decrease)  for a predetermined number of epochs, the training stops. Each dataset tested is randomly splitted into training, testing and validation. The only exception is TwiBot20, where the authors of the dataset give this split.
We evaluate the classification performances  in
terms of well-known, standard metrics, such precision, recall, F1 (the harmonic mean of precision and recall), and Matthew Correlation Coefficient (MCC) (i.e, the estimator of the correlation between the predicted class and the real
class of the samples).
The results achieved in this study are noticeable since they, in some cases, improve the state of the art. 
In general, we tried several pre-trained models, but the best results were achieved by networks based on the ResNet50 model. During the training phase, we carefully monitored the loss so as to be sure that there were no overfitting problems, and, thus, the model learned to classify correctly.


\paragraph{Comparison between state-of-art results and those by the image classification proposal for Cresci-2017}
In the paper introducing TwitBot-20~\cite{twibot20}, the authors consider two other datasets, Cresci-2017, already introduced by us above, and PAN-19\footnote{https://pan.webis.de/clef19/pan19-web/author-profiling.html}. To all 3 datasets, the authors of TwitBot-20 apply state-of-the-art detection methods, to evaluate the difference in classification performances. The best result obtained on the Cresci-2017 dataset by~\cite{twibot20} is reported in Table~\ref{tab:cresci2017result}, first row. The same table, in the second row, shows the performance results obtained by applying our method based on image classification, with ResNet50, where the loss decreases to $0.114$. From that table, we can note how our perfomance results  are equal to state-of-the-art results, with a slight improvement in MCC.


\begin{table}[h!]
\centering
\caption{Performances' comparisons on Cresci-2017: state-of-art {\it vs} image classification.}
\begin{tabular}{lcccc} 
\toprule
                & \multicolumn{4}{c}{\textbf{Cresci 2017}}  \\ 
\cmidrule{2-5}
Metric            & Accuracy & Recall & F1  & MCC        \\ 
\midrule
Feng et al.     & 0.98     & -     & 0.98     & 0.96       \\
Image classification & 0.98     & 0.98   & 0.98     & \textbf{0.98}       \\
\bottomrule
\end{tabular}
\label{tab:cresci2017result}
\end{table}


\begin{figure}
\centering
  \includegraphics[width=.7\textwidth]{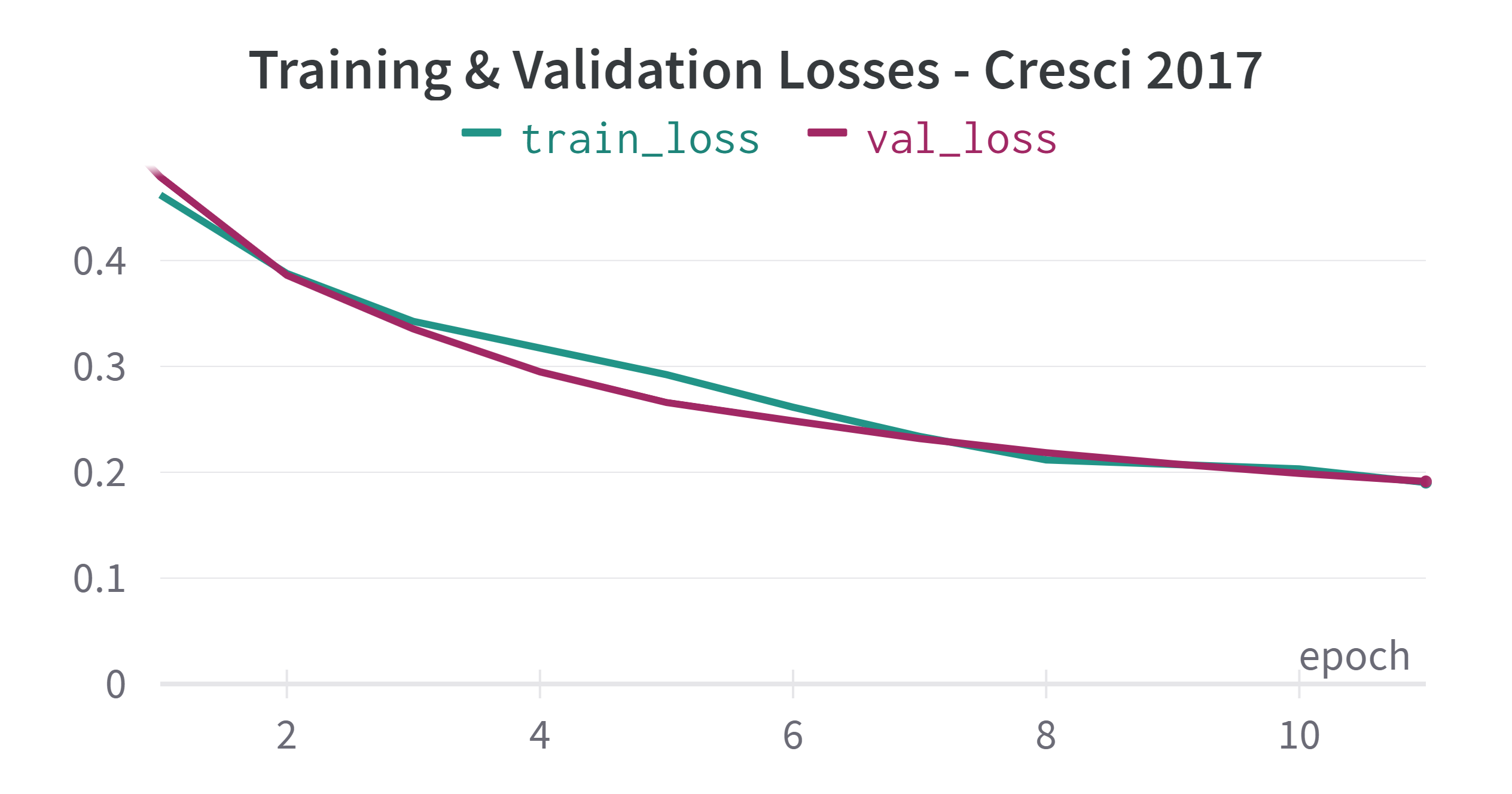}
\caption{Training and validation losses for Cresci-2017.}
\label{fig:cresci2017losses}
\end{figure}

\autoref{fig:cresci2017losses} shows the training and validation losses: the two losses have similar trends, and they decrease until they stabilize, after a number of epochs. Since the model behaves similarly in both the validation and training set, there is no overfitting~\cite{overfittingpaper}.

\paragraph{Comparison between state-of-art results and those by the image classification proposal for Cresci-Stock}

In this case, the results in \cite{articleCovid19Bots} will be taken as a reference. The comparison between the results are reported in \autoref{tab:cresci2018result}, and, as it is possible to see, the images approach improved the \textit{accuracy}, the \textit{F1 score} and the \textit{MCC}.

\begin{table}[t!]
\centering
\caption{Performances' comparisons on Cresci-Stock: state-of-art {\it vs} image classification. Results of \textit{Antenore et al.} are taken from the Table 4 of~\cite{articleCovid19Bots}.}
\begin{tabular}{lcccc} 
\toprule
                & \multicolumn{4}{c}{\textbf{Cresci stock 2018}}  \\ 
\cmidrule{2-5}
Metric            & Accuracy & Recall & F1 score & MCC        \\ 
\midrule
Antenore et al.     & 0.77     & 0.96     & 0.82     & -       \\
Image classification & \textbf{0.89}     & 0.88   & \textbf{0.89}     & \textbf{0.78}       \\
\bottomrule
\end{tabular}
\label{tab:cresci2018result}
\end{table}

\begin{figure}[ht!]
\centering
  \includegraphics[width=.7\textwidth]{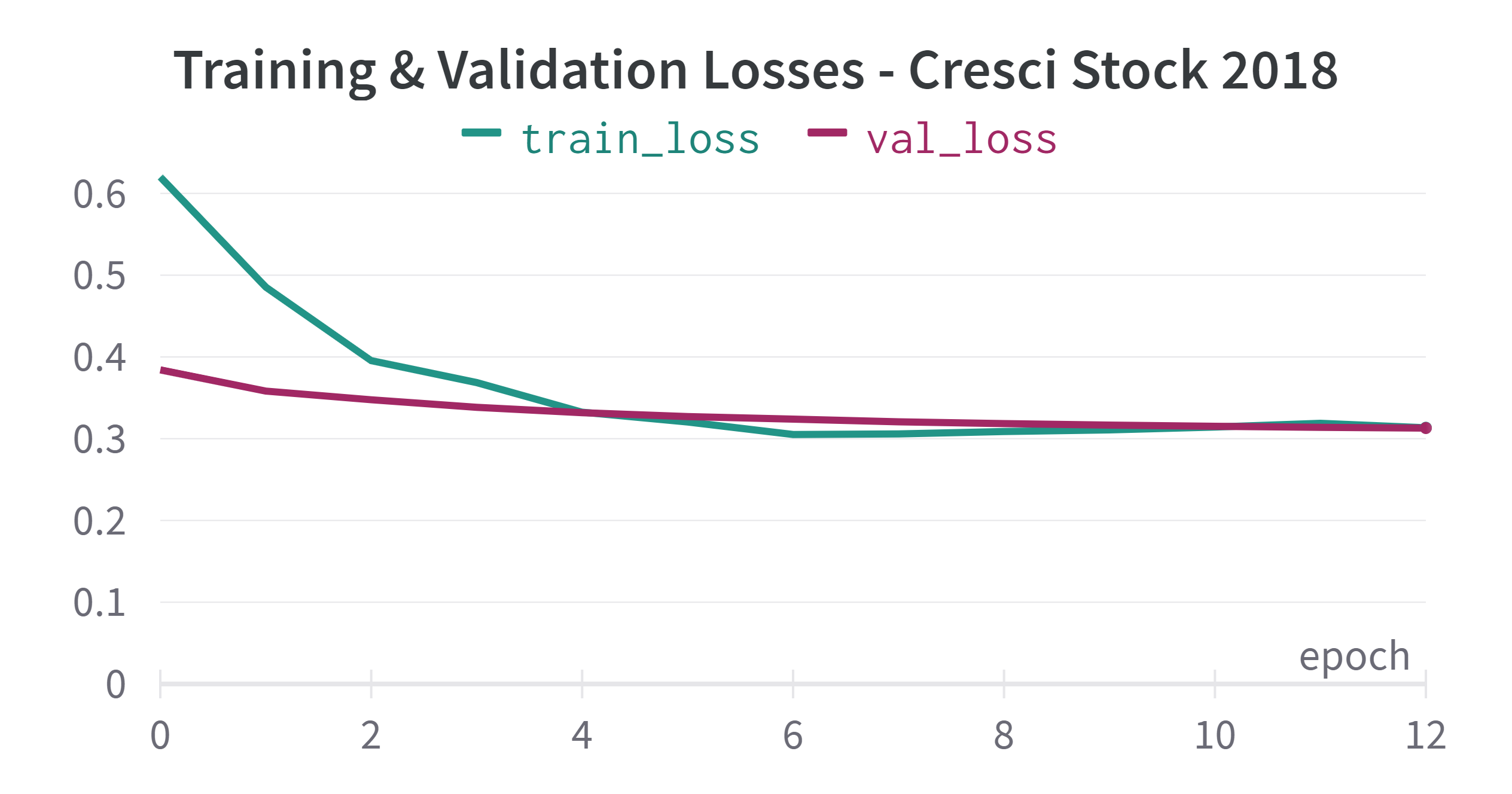}
\caption{Training and validation loss for Cresci-Stock.}
\label{fig:cresci2018losses}
\end{figure}

In \autoref{fig:cresci2018losses}, the two losses have a similar macroscopic behavior; in the training phase the loss stabilizes in fewer epochs, which is due to the larger number of samples used (60\% in training, ~30\% in validation). 
This trend rules out overfitting~\cite{overfittingpaper}. 





\paragraph{Image classification for the TwiBot20 accounts}
\begin{figure}[ht!]
\centering
\begin{subfigure}{.5\textwidth}
  \centering
  \includegraphics[width=\textwidth]{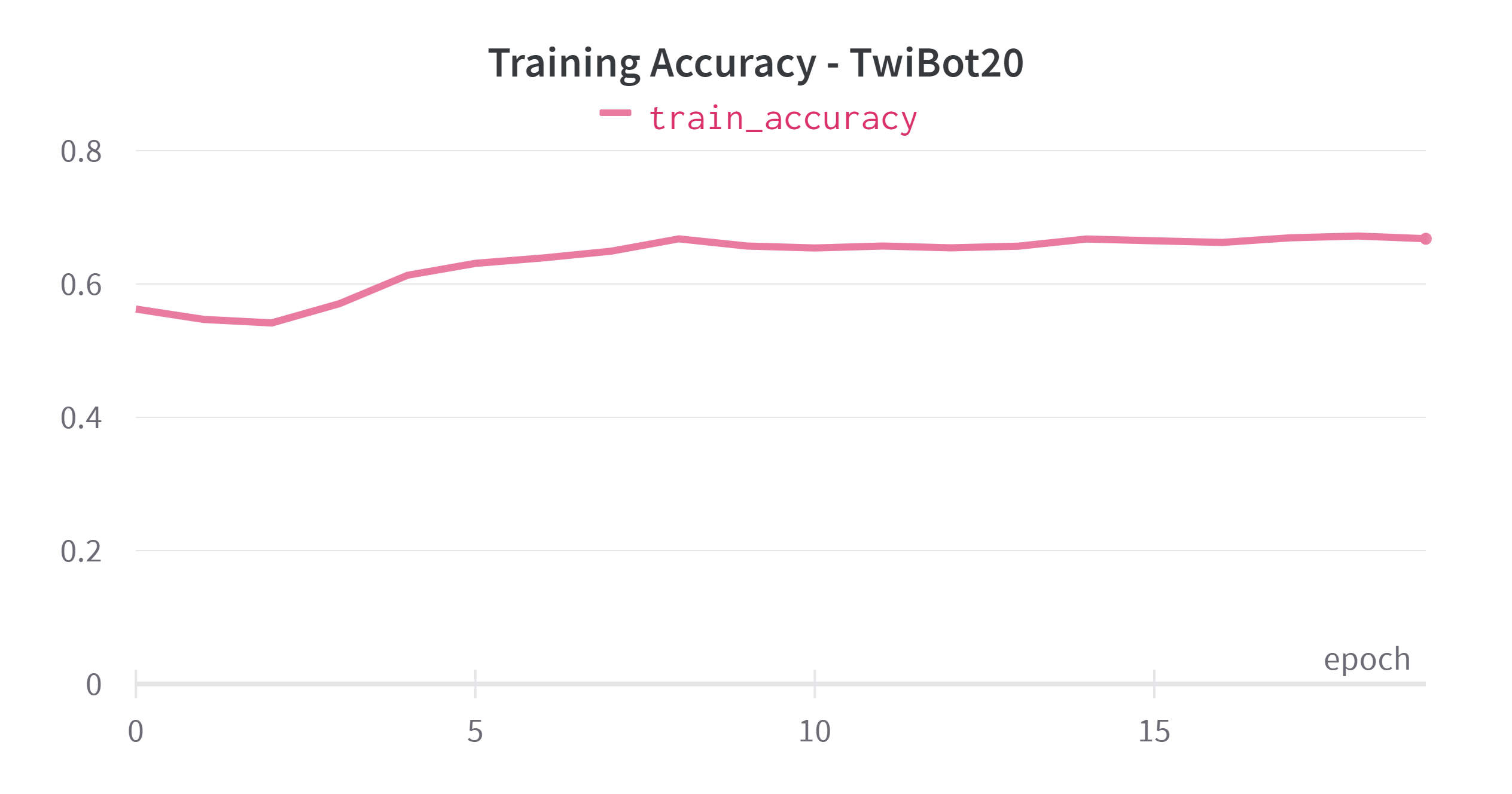}
  \caption{Training accuracy.}
  \label{fig:acctb20}
\end{subfigure}%
\begin{subfigure}{.5\textwidth}
  \centering
  \includegraphics[width=\textwidth]{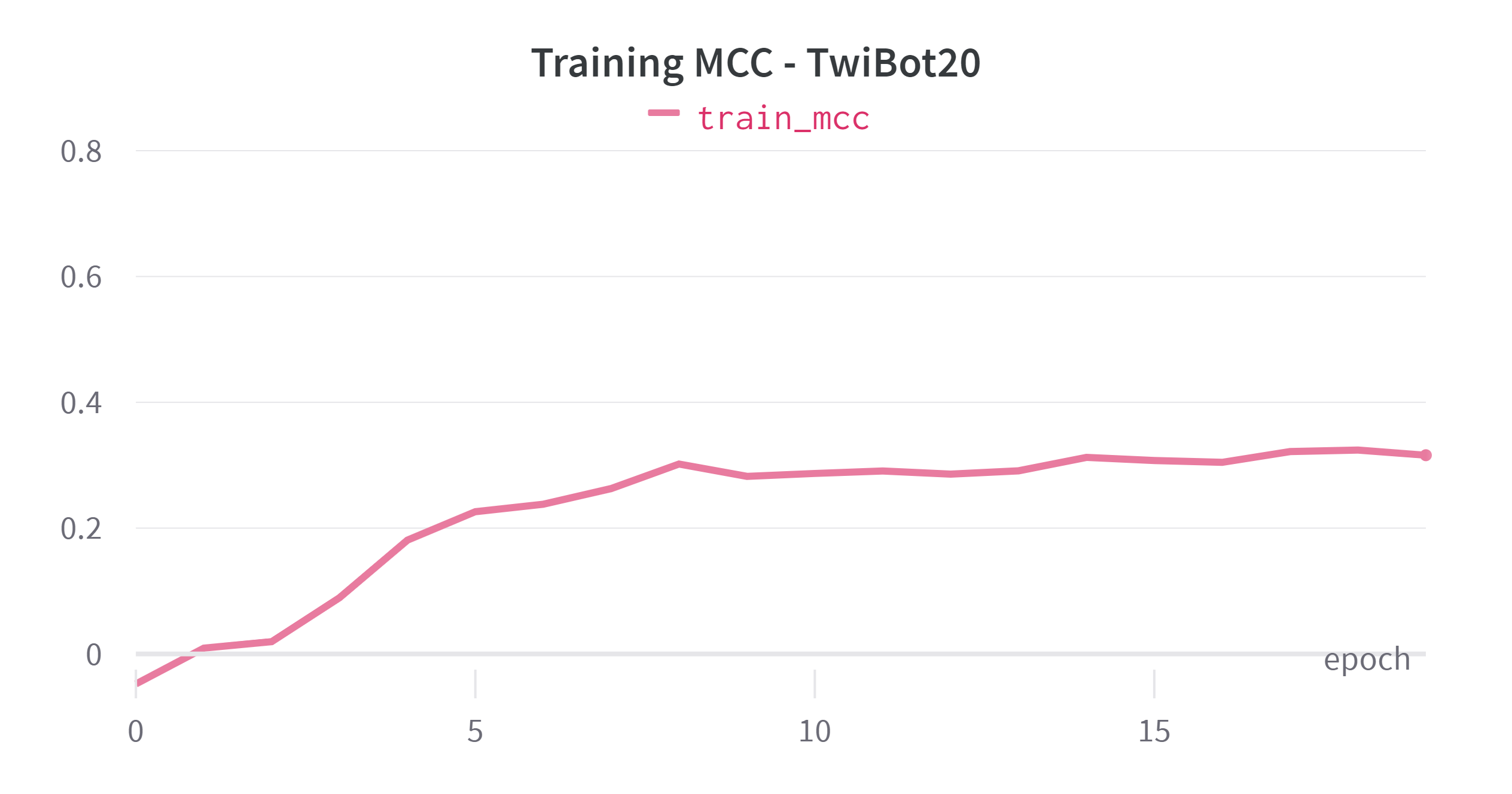}
  \caption{Training MCC.}
  \label{fig:mcctb20}
\end{subfigure}
\caption{Training accuracy and MCC considering the images originated from the TwitBot20 account timelines}
\label{fig:tb20accmcc}
\end{figure}

After applying the \autoref{alg:images_algo} to the timelines of the TwiBot20 accounts and after the CNN training phase, the classification performances are disappointing. It can be seen in \autoref{fig:mcctb20} and \autoref{fig:acctb20} in which
the accuracy does not improve so much compared to the initial phase, and MCC stabilizes between 0.3 and 0.4. The  exact numerical results are in \autoref{tab:twibot20results}, second row. 

Thus, we decide to use more account features and attach them to the user timeline. Interestingly enough, the article in~\cite{supertml}, by Sun et al., proposes an algorithm to represent tabular data as images and then proceeds with image classification. Classification achieves state-of-art results on both large and small datasets, like 
the Iris dataset\footnote{\label{supertmldatasets}\href{https://archive.ics.uci.edu/ml/datasets/iris}{Iris dataset homepage}, \href{https://www.kaggle.com/c/higgs-boson}{Higgs Boson Machine Learning challenge on Kaggle.} } and the Higgs Boson Machine Learning\footnoteref{supertmldatasets}.

Given the effective approach of~\cite{supertml}, we enlarge our feature set: the digital DNA plus all the features listed in~\autoref{tab:features_set}. 
\autoref{tab:twibot20results} shows the resulting images, for a genuine and a bot account.
The third row in~\autoref{tab:twibot20results} shows the performance results of the classification, when the image is formed with the enlarged feature set.

\begin{figure}[ht!]
\centering
\begin{subfigure}{.5\textwidth}
  \centering
  \includegraphics[scale=0.3]{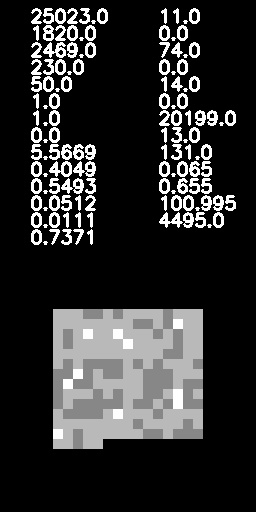}
  \caption{}
  \label{fig:hutb20}
\end{subfigure}%
\begin{subfigure}{.5\textwidth}
  \centering
  \includegraphics[scale=0.3]{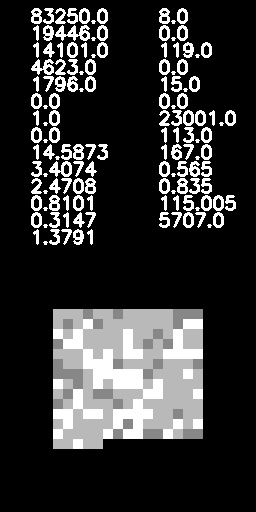}
  \caption{}
  \label{fig:bottb20}
\end{subfigure}
\caption{(a): image format for  a genuine account; (b): image format for a bot account. Both images have been created following the SuperTML algorithm, proposed in~\cite{supertml}. The upper part of the image is the list of features also reported in~\autoref{tab:features_set}. The lower part of the image is the representation of the Digital DNA.}
\label{fig:twibotsupertmlimages}
\end{figure}

\begin{table}[t!]
\centering
\caption{Performances' comparison on TwitBot20: state-of-art {\it vs} image classification {\it vs} image classification where the image is enriched with SuperTML features~\cite{supertml}.}
\begin{tabular}{lcccc} 
\toprule
                & \multicolumn{4}{c}{\textbf{TwiBot20}}  \\ 
\cmidrule{2-5}
Metric            & Accuracy & Recall & F1 score & MCC        \\ 
\midrule
Feng et al.     & 0.81     &  -    & 0.85    & 0.67      \\
Images approach & 0.67 & 0.66 & 0.61 & 0.34 \\
Images approach with SuperTML & 0.81    & 0.80   & 0.80     & 0.67       \\
\bottomrule
\end{tabular}

\label{tab:twibot20results}
\end{table}

\begin{table}[t!]
\small
\centering
\caption{List of features used.}
\begin{tabular}{p{\textwidth}} 
\hline
\multicolumn{1}{c}{\textbf{Features}}\\ 
\hline
statuses\_count, followers\_count,
friends\_count, listed\_count,
default\_profile, favourites\_count,
profile\_use\_background\_image, verified,
followers\_growth\_rate, friends\_growth\_rate,
favourites\_growth\_rate, listed\_growth\_rate,
followers\_friends\_rate, screen\_name\_length,
screen\_name\_digits\_count, description\_length,
description\_digits\_count, name\_length,
name\_digits\_count, total\_tweets\_chars\_count,
total\_urls\_in\_tweets, total\_mentions\_in\_tweets,
urls\_tweets\_rate, mentions\_tweets\_rate,
chars\_tweets\_rate, account\_age \\
\hline
\end{tabular}
\label{tab:features_set}
\end{table}

\begin{figure}[ht!]
\centering
  \centering
  \includegraphics[width=.7\linewidth]{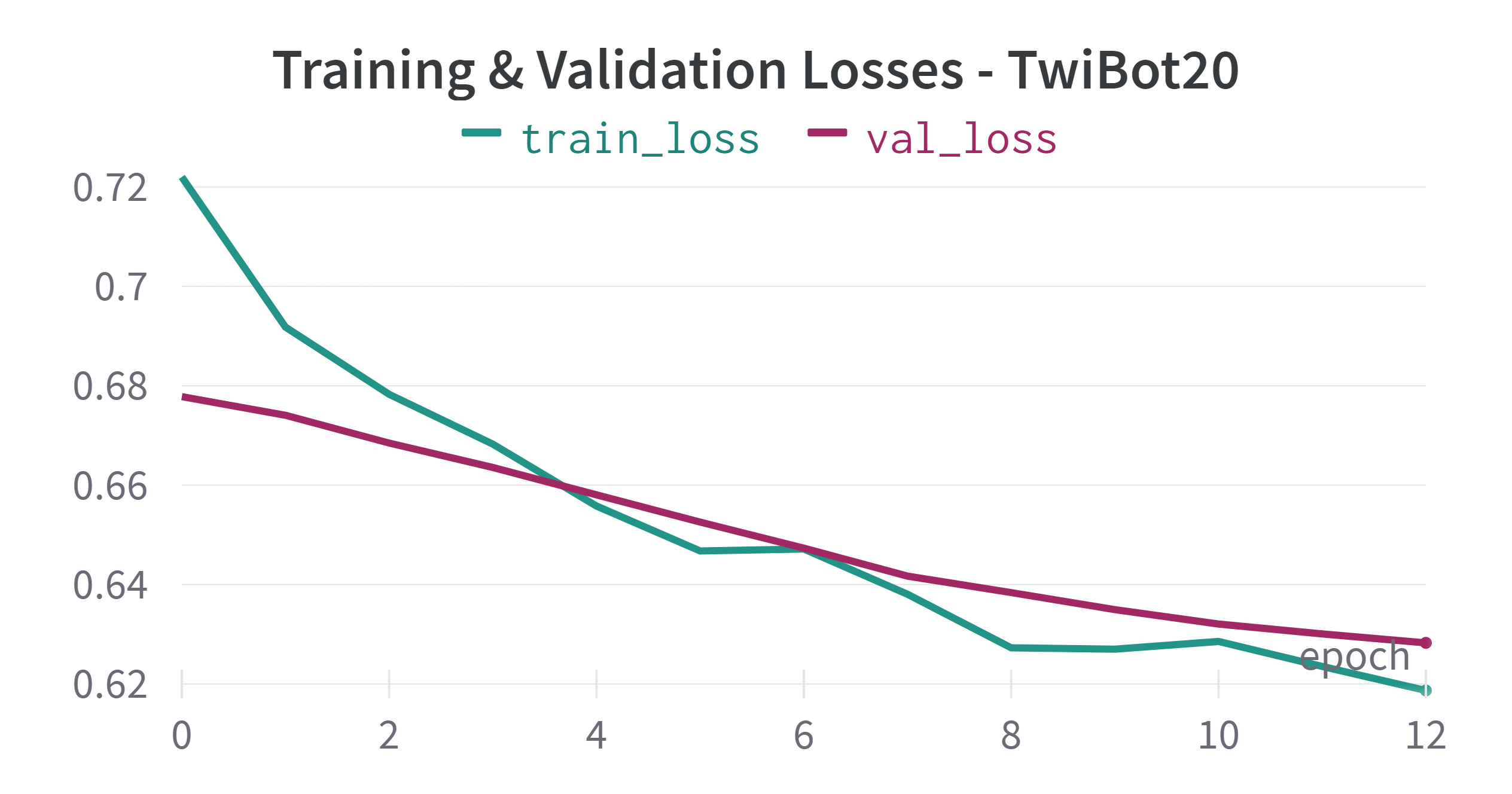}
\caption{Training and validation losses for TwitBot20 Image Classification + SuperTML}
\label{fig:tb20losses}
\end{figure}

Finally, \autoref{fig:tb20losses} shows the trends of the training and validation losses, which stabilize around $0.6$. Even in this case, the losses have very similar behavior, thus, the model is not overfitting.




\section{Conclusions}

Research in bot classification is still open, mainly due to the continuous evolution of these kind of accounts.  
This work proposed a novel method for the task, based on image classification. 
The proposed approach has been proven aligned with state-of-art results. 
Since it is tough to acquire fully representative benchmark datasets (e.g., social platforms  often block scraping, the APIs have a limited number of calls), the natural way to follow is to achieve full advantage of the available data. In the case of TwitBot20, for example, the original dataset offers numerous other pieces of information that were not exploited in the present work to obtain the images, such as, e.g., the network of interactions between accounts. As future work,  we might consider exploiting this extra information to better evaluate the proposed approach.

\bibliographystyle{splncs04}
\bibliography{bibliography}

\end{document}